\documentclass{iopart}
\usepackage{graphicx}

\begin{document}

\title{Interference of overlap-free entangled photons with a Mach-Zehnder-like interferometer}

\author{Xian-Min Jin$^{1,3}$, Cheng-Zhi Peng$^{1,2}$, Tao Yang$^1$ and Youjin Deng$^{1,3}$}
\address{$^1$ Hefei National Laboratory for Physical Sciences at Microscale and Department of Modern Physics,
University of Science and Technology of China, Hefei, Anhui, 230027,
PR China}
\address{$^2$ Physics Department, Tsinghua University, Beijing 100084, PR China}
\ead{$^3$jinxm@ustc.edu.cn; yjdeng@ustc.edu.cn}

\begin{abstract}
Using spontaneous parametric down conversion and a 50:50 beam
splitter, we generate coaxial polarization-entangled photon pairs,
of which the two photons are far separated from each other. The
photons are then sent one by one through one port of a modified
Mach-Zehnder interferometer. We observe interference fringes with a
periodicity half of the single-photon wavelength, independent of the
distance between the photons. This feature can find
applications in quantum-enhanced measurement.
\end{abstract}

\maketitle

\section{Introduction}
NOON state is a path-entangled state of $N$ photons of form
$|N0::0N\rangle=1/\sqrt{2}\; \left(|N\rangle_a |0\rangle_b+
|0\rangle_a|N\rangle_b\right)$, where $a$ and $b$ are two spatial
modes. In the past decades, NOON state has attracted much research
interest, and a variety of interferometric applications has been
found\cite{Giovannetti}, including quantum lithography\cite{Boto,
Kok2001} and Heisenberg-limited phase measurement\cite{Holland,
Bollinger, Ou1997}. The advantage of NOON state for metrology can be
understood in the concept of photonic de Broglie
wavelength\cite{Jacobson}. Within this concept, the $N$ photons in
state $|N0::0N\rangle$ are treated as a Bose-condensate ensemble,
and thus have an effective de Broglie wavelength $\lambda/N$, with
$\lambda$ the single-photon wavelength. This is analogous to the
case of a heavy massive molecule of $N$ atoms. As an application,
NOON state can be used for phase measurement which has a precision
of order $1/N$--the Heisenberg limit, beating the standard quantum
limit $1/\sqrt{N}$ that arises from statistical
fluctuations\cite{Braginsky}.

A number of experimental observations of the photonic de Broglie
wavelength $\lambda/N$ have been reported, employing double-slit
geometry\cite{Fonseca}, Mach-Zehdner (MZ)
interferometer\cite{Ou1990, Rarity, Edamatsu, Nagata}, or other
multi-particle interferometer\cite{Walther, Mitchell}. In
experiments using MZ interferometer, which is involved in most
quantum-enhanced phase measurements, Hong-Ou-Mandel (HOM)
interference\cite{Hong} is usually applied to suppress unwanted
contributions. For $N=2$, when the wave packets from the two input
ports of the MZ interferometer arrive simultaneously, the pure NOON
state $|20::02\rangle$ is generated. For $N>2$, generating a pure
NOON state by HOM interference becomes more difficult; in many
cases, even if the wave packets overlap completely, the produced
state is not a pure NOON state. For instance, the state prepared for
phase measurement in Ref.\cite{Nagata} is $\sqrt{3/4}
\;|40::04\rangle+ \sqrt{1/4} \;|22\rangle$, of which the last term
is undesired and may decrease the associated measurement precision.

In this work, we aim to circumvent the requirement that the wave
packets should at least partly overlap at the front of the MZ
interferometer to generate a NOON state. This is possible because:
1), in addition to path entanglement, one can also use other degrees
of entanglement, and 2), the de Broglie wavelength $\lambda/N$ of
NOON state does not depend on the spatial distribution of the
involved $N$ particles, as demonstrated in Ref.\cite{Walther}.

\section{Experimental Set-up}

The setup of our experiment is schematically shown in Fig. 1. It
contains two parts: the photon-source generator (Fig. 1a) which
produces polarization-entangled photon pairs and aligns the photons
into a co-axis, and a modified MZ interferometer (Fig. 1b).

\begin{figure}
\begin{center}
\includegraphics[width=0.97\linewidth]{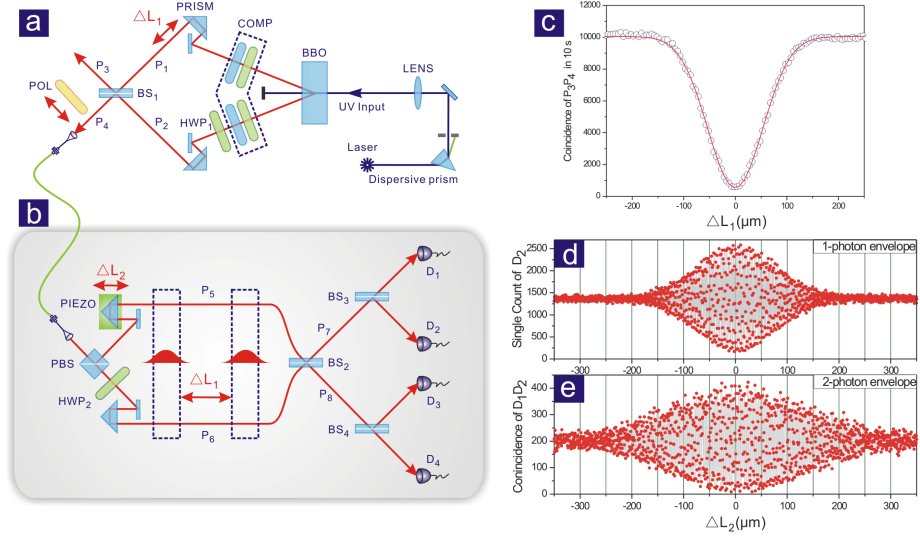}
\end{center}
\caption{Schematic experimental setup and interference searching.
\textbf{(a)}. Preparation of coaxial polarization-entangled state
$|\Phi^{+}\rangle_{P4}$ by Eq.~(\ref{Eq.2}). Polarization
entanglement comes from the process of spontaneous parametric down
conversion by pumping a nonlinear crystal. BS1 serves as a optical
mixer coupling entangled photons to path P4. $\Delta L_{1}$ is tuned
by adjusting micrometer-resolution prism. \textbf{(b)}. A modified
Mach-Zehdner interferometer. \textbf{(c)}. The curve for HOM
interference at BS1, which displays two-photon coincidence rate
versus the relative optical delay between the interfering photons
$\Delta L_1$. \textbf{(d), (e)}. one- and two-photon interference
envelope. BBO: $\beta$-barium-borate crystal; HWP: half wave plate;
COMP:HWP and BBO(1mm); POL: polarization plate; BS: beam splitter;
PBS: polarization beam splitter; PIEZO: piezo ceramics}\label{Fig.1}
\end{figure}

In Fig. 1a, a semiconductor blue laser beam (power 34.5$mw$, waist
100$\mu m$, and central wavelength 405$nm$) is incident on a
2$mm$-barium-borate (BBO) crystal. As a consequence, pairs of
polarization-entangled photons are generated via type-II
SPDC\cite{Kwiat}, of which the central wavelength is 810$nm$. The
down-converted extraordinary and ordinary photons have different
propagating velocities, and travel along different paths inside the
crystal due to birefringent effect of the BBO crystal. The resulting
walk-off effects are then compensated by a combination of a half
wave plate (HWP) and an additional 1 $mm$ BBO crystal in each arm.
Photons are collected by single mode fibers (SMF). The count rate is
about 80$k/s$ in each arm, and the coincidence is about 20$k$ pairs
per second. As a result, we prepare the polarization entanglement as

\begin{eqnarray}
|\Phi^{+}\rangle_{s}=1/\sqrt{2} \; \left(|H\rangle_{1}|H\rangle_{2}+
|V\rangle_{1}|V\rangle_{2} \right) \; , \label{Eq.1}
\end{eqnarray}

where where $|H \rangle (|V\rangle)$ denotes horizontal (vertical)
polarization, the subscripts $1$ and $2$ specify spatial modes, and
subscript $s$ means state of source. The visibilities for the
polarization correlations are about 98.1$\%$ for $|H \rangle
/|V\rangle$ basis and 92.6$\%$ for
$|+45^{\circ}\rangle/|-45^{\circ}\rangle$ basis, without the help of
narrow bandwidth interference filters. This implies that we have
produced a high-quality polarization-entangled photonic source. The
photons from pathes P1 and P2 are combined at a 50:50 beam splitter
(BS). A prism-built-in manual translation stage is placed on path P1
to adjust the path difference $\Delta L_1$, of which the precision
is of order $\mu m$. The photons along path P4 are then guided into
one port of the modified MZ interferometer (Fig. 1b).

The photon state on path P4 is usually a mixed state, depending on
the path difference $\Delta L_1$. When $\Delta L_1 \geq \xi $ with
$\xi$ the single-photon coherent length, the wave packets do not
overlap at BS1 and each photon has the same probability to transmit
or be reflected. Thus, the photon state on P4 can be a single-photon
state $H$ and $V$ with probability $1/4$, respectively, or the
maximally polarization-entangled state with probability $1/4$. The
entanglement is also at the temporal mode, let $t_1 \leq t_2 $
specify the time that the two photons arrive at the polarization
beam splitter (PBS) of the modified MZ interferometer (Fig. 1b), and
the state can be written as
\begin{eqnarray}
|\Phi^{+}\rangle_{P4}=1/\sqrt{2} \;
\left(|H\rangle_{t1}|H\rangle_{t2}+|V\rangle_{t1}|V\rangle_{t2}
\right)
\; ,
\label{Eq.2}
\end{eqnarray}
where subscript $P4$ refers to path P4. When interval $\Delta L_1 <
\xi $, the HOM interference occurs. In particular, when $\Delta L_1
=0$ i.e., the two wave packets completely overlap at BS1, the state
that a photon is on path P3 and the other on P4 is eliminated, and
the state on P4 is a pure state described by Eq.~(\ref{Eq.2}).

The HOM interference curve is obtained by scanning the
position of prism and measuring the coincidence at two output modes
$P3$ and $P4$, as shown in Fig. 1c. The visibility
is determined as $V_{HOM}=(C_{plat}-C_{dip})/C_{plat}=(94.5\pm 0.4)\%$,
where $C_{plat}$ is the noncorrelated coincidence rate at the plateau and $C_{dip}$
is the interfering coincidence rate at the dip.
Such a high visibility suggests that we have established accurate
spatial and temporal overlap for the two entangled photons.
Further, the interference curve in Fig. 1c can be approximately described
by a Gaussian function, which yields the single-photon coherent
length $\xi \approx 126 \mu m$.

The setup in Fig. 1b is obtained by replacing the front BS of the
standard MZ interferometer by a PBS and a HWP set at $45^{\circ }$.
A manual translation stage with micrometer precision and a piezo
ceramics of precision 1$nm$ is placed on path P5 to adjust the path
difference $\Delta L_2$. The two pathes are combined again at BS2.
Additional two BSs are placed on pathes P7 and P8, respectively, and
avalanche photo diodes are used for photon detection.

Figure 1b can be regarded as a modified MZ interferometer. Consider
a single-photon state $|+45^{\circ}\rangle= 1/\sqrt{2} \; (|H
\rangle + |V \rangle)$. The horizontally polarized part $H$ will
transmit through the PBS onto path 6, and while the vertically
polarized part will be reflected onto path 5. Thus, after the PBS,
the state is $1/\sqrt{2} \;  (|V \rangle_5 + |H \rangle_6)$. Then,
the HWP will erase  polarization information by rotating $H$ to $V$
at path 6. Taking into account the phase difference due to interval
$\Delta L_2$, the photon state becomes $1/\sqrt{2} \; (|1 \rangle_5
|0 \rangle_6 + e^{i\phi} |0 \rangle_5 |1 \rangle_6)$. Similar
analysis yields that a $N$-photon Greenberger-Horne-Zeilinger (GHZ)
state\cite{Bouwmeester} $1/\sqrt{2} ( |H \rangle_{1} |H \rangle_{2}
\cdots |H \rangle_{N} + |V \rangle_{1} |V \rangle_{2} \cdots |V
\rangle_{N}$ on path 4 will be transferred into a state immediately
before BS2 as $1/\sqrt{2} \; (|N \rangle_5 |0 \rangle_6 + e^{iN\phi}
|0 \rangle_5 |N \rangle_6)$. Clearly, such an action for $N=1$ and
$N=2$ is analogous to previous experiments in a standard MZ
interferometer.

\section{Experimental results}

We carried out a few experiments, for a single-photon
self-interference and the interference of the polarization-entangled
pairs with various $\Delta L_1$, respectively. Some of these
experiments are mainly for testing purposes and the others are to
demonstrate the interference of overlap-free entangled photons

\subsection{Single-photon interference}

In this experiment, the BS1 in Fig. 1a is replaced by a polarization
plate (POL) oriented at $+45^{\circ}$, and thus the photon state on
path P4 becomes the linear polarization state
$|+45^{\circ}\rangle=1/\sqrt{2}(|H\rangle+|V\rangle)$. As described
earlier, the state before BS2 is $1/\sqrt{2} \; (|1 \rangle_5 |0
\rangle_6 + e^{i\phi} |0 \rangle_5 |1 \rangle_6)$. After interfering
at BS2, ideally, the probability to detect the single photon on path
P7 is $P_{7}=(1+\cos \phi )/2$, with $\phi = \Delta L_2 /\lambda$,
and the probability on path P8 is $P_{8}=(1- \cos \phi )/2$.

We first scanned $\Delta L_{2}$ in a large range and obtain an
interference envelope shown in Fig. 1d. This envelope was actually
used to calibrate the parameter $\Delta L_{2}$: $\Delta L_{2}=0$
corresponds to the position at the middle of the interference
envelope where the interference amplitude is maximal. Further, the
single-photon coherent length $\xi$ can be obtained as the Full
width at half maximum (FWHM) of the envelope, which is about 130$\mu
m$, consistent with the result deduced from the HOM interference
(Fig 1.c).

We then set $\Delta L_2 \approx 0$ and scanned the piezo ceramics
with a step of 10nm. Indeed, the counting rates on Path P7 and P8
were observed to oscillate with periodicity $\lambda =810 nm$, as
shown in Fig. 2a. The curves are of high quality and well described
by $[1 \pm \cos (\Delta L_2/\lambda)]/2$.

\begin{figure}
\begin{center}
\includegraphics[width=0.97\linewidth]{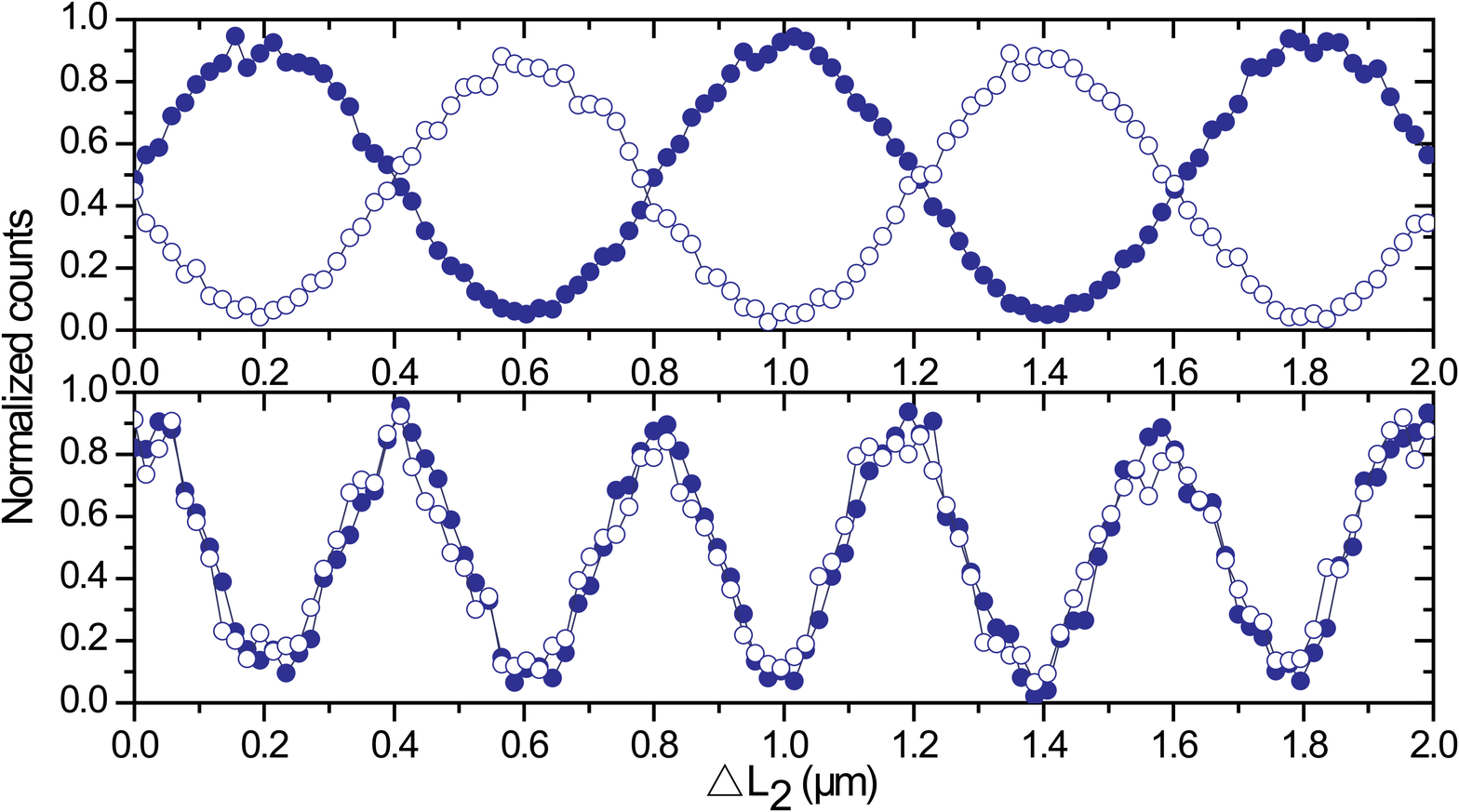}
\end{center}
\caption{Characteristics of one- and two-photon interference.
\textbf{(a)}. one-photon count rate at path P7 (blue solid circles)
and at P8 (blue open circles) as the interval $\Delta L_2$ is
varied. \textbf{(b)}. Two-photon coincidental detection rates of
$D_1D_2$ (blue solid circles) and $D_3D_4$ (blue open circles)
versus $\Delta L_{2}$.} \label{Fig.2}
\end{figure}

\subsection{Interference of entangled photon pairs with $\Delta L_1 =0$}

With interval $\Delta L_1 =0$, the wave packets of the entangled
photons arrive simultaneously at BS1 and the PBS, and the state
immediately before BS2 is
$1/\sqrt{2}(|2\rangle_{5}|0\rangle_{6}+e^{i2\phi}|0\rangle_{5}|2\rangle_{6})$.
After BS2, the state becomes $1/\sqrt{8} \;
((1-e^{i 2 \phi}) |2 \rangle_7 |0\rangle_8
- (1-e^{i 2 \phi}) |0 \rangle_7 |2\rangle_8 -
\sqrt{2} i ( 1+e^{i 2 \phi}) |1 \rangle_7 |1 \rangle_8)$.
Therefore, the probability for both photons on path P7 or P8
is $P_7=P_8=(1-\cos (2\phi))/4$, and the probability for one photon on P7 and the
other on P8 is $(1+\cos (2\phi))/2$. These probabilities are measured by
the coincidence rate of $D_1 D_2$, $D_3 D_4$, and $D_2 D_3$, respectively.

We first varied $\Delta L_2$ within a big range to determine the
interference envelope, shown in Fig. 1e. Clearly, this envelope is
much wider than that for the single-photon self-interference in Fig.
1d. The FWHM of the interference envelope in Fig. 1e is about 300
$\mu m$. To our knowledge, this is first determination  of the full
interference envelope of a photonic de Broglie wave.

The interference fringes of the entangled photon pairs is obtained
by varying $\Delta L_2$ near the middle of the envelope with a step
of 10$nm$. The coincidence rates of $D_1 D_2$ and $D_3 D_4$ are
shown in Fig. 2b. Indeed, the oscillation periodicity is $\lambda /2
\approx 400 nm$. The two oscillation curves also have the same
phase, as predicted earlier. This confirms the concept of photonic
de Broglie wave for NOON state.

\subsection{Interference of entangled photon pairs
with $\Delta L_1 \neq 0$}

To demonstrate how the two-photon interference looks like for
$\Delta L_1 \neq 0$, we theoretically consider the limiting case
$\Delta L_1 \gg 0$ such that one photon in the entangled pair
described by Eq.~(\ref{Eq.2}) has already been detected while the
other one is still before the PBS in Fig. 1b. Simple analysis yields
that the state is
\begin{itemize}
\item immediately before BS2:
$1/\sqrt{2}\; \left(|H\rangle_{t2} |1 \rangle_{6,t1}
+e^{i \phi}|V\rangle_{t2,}|1\rangle_{5,t1}\right)$,
\item immediately after BS2:
$1/2\; \left[\left(|H\rangle_{t2} -e^{i\phi} |V\rangle_{t2} \right) |1 \rangle_{7,t1}
+\left(|H\rangle_{t2} +e^{i\phi} |V \rangle_{t2}\right) |1\rangle_{8,t1}\right]$.
\end{itemize}
Therefore, with equal probability $1/2$, the first photon will
be detected on path P7 or P8. Accordingly, the wave packet collapses onto
$1/\sqrt{2}(|H\rangle_{t2} \pm e^{i\phi} |V\rangle_{t2} )$,
where $\pm$ represents the two cases, respectively. Some calculations
give that the state of the second photon immediately after BS2 is
\begin{equation}
1/2\; \left[\left(1\mp e^{i2 \phi}\right) |1 \rangle_{7,t2} +
\left(1 \pm e^{i2\phi}\right) |1\rangle_{8,t2}\right] \; .
\end{equation}
This means that the probability for both the first and the second photon
travelling along the same path (either P7 or P8) is $P_7=P_8=(1+\cos 2 \phi )/2$, and
that the two photons travel along the different pathes with probability
$(1-\cos 2 \phi )/2$. This is the same as the case $\Delta L_1 =0$.

In short, it is plausible to expect that the coincidence rate
$\propto 1+\cos 2 \phi $ of $D_1 D_2$ and $D_3 D_4$ does not depend
on the interval $\Delta L_1$. In the present experiment setup, the
only effect of $\Delta L_1$ is to affect the efficiency of preparing
the state by Eq.~(\ref{Eq.2}). In addition to $\Delta L_{1}=0$, we
carried out experiments for $\Delta L_{1}=200 \mu m$ and 10000$\mu
m$. The interval $\Delta L_{1}=200 \mu m$ is larger than the
single-photon coherent length $\xi \approx 126 \mu m$ but still
within the coherent length of the pump laser $\xi_{\rm pump} \approx
300 \mu m$, while $\Delta L_{1}=1000 \mu m$ is significantly larger
than $\xi_{\rm pump}$.

Figure 3 shows the coincidence rates for $D_1D_2$ as a function of
$\Delta L_{2}$. In all the three cases, the periodicity is always
about 400$nm$, half of the single-photon wavelength $\lambda =810
nm$. This confirms our expectation mentioned above. Nevertheless,
the amplitude of the coincident events decreases as $\Delta L_{1}$
becomes larger. For $\Delta L_{1} < \xi $, HOM interference occurs,
which improves the efficiency of the preparation of state by
Eq.~(\ref{Eq.2}). For $\Delta L_{1} > \xi $, the coupling efficiency
to SMF becomes dominant. In particular, the displacement $\Delta
L_{1}=10000\mu m$ significantly exceeds the Rayleigh length defined
by our SMF, and thus leads to the suppression of the coincidence
rate without readjustment. To quantify the quality of the
interference, we define
\begin{eqnarray}
V=(C_{max}-C_{min})/(C_{max}+C_{min}) \; ,
\label{Eq.6}
\end{eqnarray}
where $C_{max} (C_{min})$ is the correlated coincidence rate at the
peak (valley) of the two-fold coincidence curve. In our experiment,
$V$ values calculated from sinusoidal interference fringes are
$0.98\pm0.01$, $0.95\pm0.03$, $0.98\pm0.03$ respectively. This
suggests that the interference fringes in Fig. 3 are all of high
quality.

\begin{figure}
\begin{center}
\includegraphics[width=0.97\linewidth]{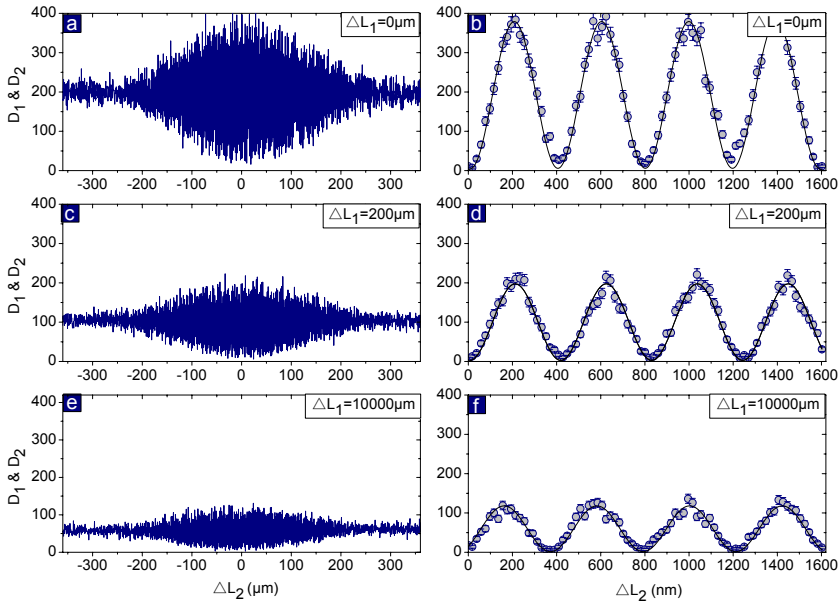}
\end{center}
\caption{Two-photon interference envelope and fringes versus
interval $\Delta L_2$. Error bars are given by Poissonian
statistics. \textbf{(a)(b)}. $\Delta L_{1}=0 \mu m$.
\textbf{(c)(d)}. $\Delta L_{1}=200 \mu m$. \textbf{(e)(f)}.  $\Delta
L_{1}=1000 \mu m$.} \label{Fig.3}
\end{figure}

\section{Discussion}

This experiment was finished in May 2008 when one of us (XJ) was
working on Ref.\cite{Jin}. Thanks to Refs.\cite{Walther}, we become
aware that our experimental results are of significant scientific
values. Meanwhile, we also realized that our experiment could have
been further improved. 1), the function of BS1 is just like an
optical mixer. In principle, it can be replaced by an optical switch
or a WDM (wavelength-division multiplexing). Accordingly, the
efficiency of generating the state by Eq.~(\ref{Eq.2}) can be
optimal $p=1$. 2), one could have tried to increase the interval
$\Delta L_1$ further to mimic the limiting case $\Delta L_1 >>0$,
discussed at the beginning of Section 3.3.

It is already known that the coherent length of a NOON state is
normally much larger than the associated single-photon coherent
length. This is demonstrated by the interference envelope in Fig.
1e, which is for $\Delta L_1 =0$ where the two photons are not
distinguishable. Nevertheless, for $\Delta L_1 >> \xi$, the wave
packets of the two photons are far apart from each other, and thus
they can be distinguished by their spatial positions. For instance,
the interference at BS2 seems to depend on the overlap of the
single-photon wave packets from pathes P5 and P6. In this case, it
is not clear (at least to us) whether the coherent length is still
much larger than the single-photon coherent length. Therefore, we
also measured the whole interference envelopes for $\Delta L_1=0$,
200$\mu m$, and 1000$\mu m$. There are shown in Fig. 3, which
suggests that, like the interference periodicity, the coherent
length of the entangled pair does not depend on the interval $\Delta
L_1$.

Finally, we mention that the present experimental setup can be
equally well used for the $N$-photon GHZ state\cite{Bouwmeester}
$1/\sqrt{2} ( |H \rangle_{1} |H \rangle_{2} \cdots |H \rangle_{3} +
|V \rangle_{1} |V \rangle_{2} \cdots |V \rangle_{N}$. The
periodicity of the associated interference will become $\lambda/N$,
and thus can be used for quantum-enhanced measurement. In comparison
with the conventional MZ interferometer, it alleviates the
requirement that the involved photons should arrive simultaneously
at the first BS. Actually, it also provides a way to generate a NOON
state from a GHZ state.

\section*{Acknowledgements}
We are grateful for insightful discussions with Y. Yamamoto. This
work was also supported by the National Natural Science Foundation
of China and China Postdoctoral Science Foundation.

\end{document}